\documentclass[conference]{IEEEtran}
\usepackage{cite}
\usepackage{amsmath,amssymb,amsfonts}
\usepackage{algorithm}
\usepackage{algpseudocode}
\usepackage{graphicx}
\usepackage{textcomp}
\usepackage{xcolor}
\usepackage{array}
\usepackage{tabularx}
\usepackage{bbding}
\usepackage{float}
\usepackage{subcaption}
\usepackage{caption}
\usepackage{booktabs}
\usepackage{makecell}
\usepackage[hidelinks]{hyperref}
\usepackage{tikz}

\begin{document}
\title{LLM-based Continuous Intrusion Detection Framework for Next-Gen Networks}

\author{
    \IEEEauthorblockN{Frédéric Adjewa\IEEEauthorrefmark{1}\IEEEauthorrefmark{2}, Moez Esseghir\IEEEauthorrefmark{1}, Leïla Merghem-Boulahia\IEEEauthorrefmark{1}}
    \IEEEauthorblockA{\IEEEauthorrefmark{1} University of Technology of Troyes, LIST3N-ERA, France}
    \IEEEauthorblockA{\IEEEauthorrefmark{2} Computer Science and Information Systems departement, Institut Universitaire Saint Jean du Cameroun, Yaounde, Cameroon}
    \IEEEauthorblockA{Email: \{frederic.adjewa, moez.esseghir, leila.merghem\_boulahia\}@utt.fr}
}

\maketitle

\begin{center}
    \textbf{This work has been submitted to the IEEE for possible publication. Copyright may be transferred without notice, after which this version may no longer be accessible.}
\end{center}

\begin{abstract}
In this paper, we present an adaptive framework designed for the continuous detection, identification and classification of emerging attacks in network traffic. The framework employs a transformer encoder architecture, which captures hidden patterns in a bidirectional manner to differentiate between malicious and legitimate traffic. Initially, the framework focuses on the accurate detection of malicious activities, achieving a perfect recall of 100\% in distinguishing between attack and benign traffic.
Subsequently, the system incrementally identifies unknown attack types by leveraging a Gaussian Mixture Model (GMM) to cluster features derived from high-dimensional BERT embeddings. This approach allows the framework to dynamically adjust its identification capabilities as new attack clusters are discovered, maintaining high detection accuracy. Even after integrating additional unknown attack clusters, the framework continues to perform at a high level, achieving 95.6\% in both classification accuracy and recall.
The results demonstrate the effectiveness of the proposed framework in adapting to evolving threats while maintaining high accuracy in both detection and identification tasks. Our ultimate goal is to develop a scalable, real-time intrusion detection system that can continuously evolve with the ever-changing network threat landscape.
\end{abstract}


\section{Introduction}
Network systems are becoming increasingly complex due to ongoing technological advancements. 
Next generations of network are expected to enable various interactions and exchanges over the Internet, supporting services such as digital brain-computer interfaces, in-body health networks, and extended reality. This transition will be underpinned by achieving data rates around terabits per second with ultra-low latency.
This continuous evolution and expansion of network infrastructure lead to an extremely heterogeneous environment characterized by massive volumes of data. This can complicate network monitoring, exposing systems to malicious activities. Furthermore, as new services emerge from network improvements, sophisticated threats are likely to increase correspondingly, becoming harder to detect \cite{Diaf}. Therefore, it is imperative to develop robust security measures within these highly heterogeneous ecosystems.
Various machine learning methods have been explored to enhance security across diverse network environments, including smart grids \cite{sahani2023machine}, virtual private networks \cite{parenreng2023network}, and IoT ecosystems \cite{ravi2023deep}. Among these methods, Support Vector Machines (SVMs) are widely utilized for their ability to classify data into distinct categories, making them effective for detecting intrusions and anomalies within network traffic. Boosting methods enhance the performance of weak classifiers by combining them into a strong ensemble classifier, thereby improving detection rates for malicious activities. Deep Neural Networks (DNNs) leverage multiple layers of neurons to learn intricate patterns in data, which proves particularly useful for identifying advanced threats in large-scale networks. Recurrent Neural Networks (RNNs) are specifically designed to analyze sequences of network packets, enabling the identification of patterns indicative of cyber threats. Long Short-Term Memory (LSTM) networks excel at detecting temporal patterns in network traffic and are frequently employed in applications such as detecting Distributed Denial of Service (DDoS) attacks \cite{ferrag2020deep}.
In addition to their diversity, the data in these ecosystems are dynamic, presenting the challenge of encountering unknown data flows during analysis. Machine learning-empowered intrusion detection systems (IDSs) often struggle to keep pace with the constant evolution of network flows, as they are typically trained to detect a fixed or predefined set of attacks \cite{10547921}. However, in real-world scenarios, intrusion data are collected incrementally. As new data emerge, these models may suffer from catastrophic forgetting, a prevalent issue in machine learning where previously learned classes are forgotten when training on new ones \cite{data2021t}.
Recently, techniques involving LLMs have demonstrated an ability to effectively manage vast volumes of data while understanding the underlying context within them. Given the unknown pattern in next generation of networks, leveraging LLMs presents a promising approach for efficiently processing and analyzing such data. Studies have shown that LLMs can address various challenges in telecommunications, such as optimizing the reward process in reinforcement learning \cite{booth2023perils}, and providing zero-shot image classification in complex signal transmissions \cite{zhou2024large}.
However, it is important to note that LLMs were not initially designed for these applications; their primary utility lies in natural language processing (NLP). Consequently, directly applying them to domain-specific tasks can be challenging, necessitating fine-tuning for optimal performance. To the best of our knowledge, this work proposes the first hybrid incremental intrusionTo the best our knowledge, this work proposes the first hybrid continuous intrusion detection framework that utilizes LLMs to address emerging threats. The code of our simulation is available \footnote{\url{https://github.com/freddyAdjh/LLM-CIDF-next_gen}}.

Our contributions in this work are as follows:
\begin{itemize}
    \item We present a novel framework that leverages language models to detect and identify malicious traffic, effectively capturing contextual information with high accuracy.
    \item We propose an adaptive clustering approach based on a Gaussian Mixture Model applied to network flow embeddings, enabling the dynamic identification and categorization of previously unknown attack types for continuous learning and model updating.
    \item We demonstrate the framework's effectiveness in adapting to emerging threat patterns, achieving high accuracy and recall in detection and identification.
\end{itemize}

\section{Related Work}

Considered as the first line of defense for network systems, Network Intrusion Detection Systems (NIDS) monitor logs and traffic to detect intrusion attempts, assisting in decision-making when necessary. NIDS are typically deployed alongside other security measures, such as encryption techniques and authentication mechanisms. In recent years, machine learning (ML) techniques have been increasingly employed to manage the complexity of traffic flow, demonstrating their potential to enhance IDS concepts. Numerous studies have focused on refining these systems to maintain effectiveness in the face of the constantly evolving diversity of cyber-attacks.

For instance, \cite{basumallik2019packet} developed a CNN-based data filter to detect false data injection attacks (FDIA) in power systems using time-series PMU data. Similarly, \cite{bocu2022real} proposed a machine learning-based IDS that utilizes a Convolutional Neural Network (CNN) for real-time protection and prevention in 5G networks. Although this system exhibited fast attack detection, its low accuracy, as indicated by a high false negative rate, presents a significant challenge for intrusion detection.

In a related work, \cite{abusitta2019deep} suggested the use of a denoising autoencoder for proactive intrusion detection based on noisy feedback. This solution is particularly valuable as it can leverage incomplete data, addressing the real-world challenge of obtaining comprehensive datasets. Furthermore, \cite{articleSydney} proposed an IDS based on a Feed Forward Deep Neural Network (FFDNN) combined with a feature selection algorithm. Their model outperformed traditional machine learning methods in accuracy when evaluated on the NSL-KDD dataset. Likewise, \cite{yin2017deep} introduced an RNN-based intrusion detection system, which demonstrated superior performance in both binary and multiclass classifications compared to traditional methods. Their study also explored the effects of varying neuron count and learning rate on model accuracy, using a benchmark dataset.
To address DDoS attacks in heterogeneous IoT networks, \cite{10014866} presented a smart LSTM-based IDS, achieving high accuracy while maintaining a lightweight and less complex architecture compared to state-of-the-art deep learning approaches.
While the application of these machine learning techniques aims to create more resilient and adaptive security solutions in the face of evolving network vulnerabilities, many approaches struggle with issues such as catastrophic forgetting and evolving detection. For example, \cite{alaei2017incremental} introduced a solution based on the Naive Bayes model, which is inherently incremental and thus better suited to handle the continual arrival of packet flows. However, the assumption of data independence a key aspect of Bayesian models limits the effectiveness of this approach, and the performance obtained was suboptimal.

\cite{data2021t} explored the use of deep neural networks to learn and retain new knowledge. However, the expansion of the model with new data often leads to uncontrolled growth, as new data tends to be plentiful. \cite{rebuffi2017icarl} highlighted that for a model to be considered incremental, it must maintain a stable memory footprint or grow minimally. 

Recent advances in LLMs have shown significant potential in efficiently processing natural language. In our prior work \cite{adjewa2024efficient}, we demonstrated that it is feasible to deploy an LLM on resource-constrained devices in a decentralized context while maintaining good overall accuracy.

However, BERT's static training on a fixed set of attacks introduces uncertainty when new attack flows emerge, as it tends to classify novel attacks based on patterns from previously known categories. Similarly, \cite{Diaf} proposed a proactive network intrusion prediction framework that integrates fine-tuned GPT and BERT models with an LSTM classifier to predict and evaluate IoT network traffic. While effective, this framework primarily addresses the binary nature of network traffic.
In an effort to enhance security in 6G architectures, \cite{sedjelmaci2023enabling} suggested combining zero-trust architecture with artificial intelligence by deploying AI-based agents to protect the network from internal, external, and zero-day attacks. However, their approach lacks a focus on detailed analysis of newly detected threats and does not utilize advanced AI models for in-depth flow analysis, an essential feature as threats become increasingly sophisticated.

\begin{table*}[ht!]
\centering
\caption{Comparative Analysis of Related Work and Proposed Framework}
\begin{tabular}{|p{2cm}|p{3cm}|p{3cm}|p{3cm}|p{4cm}|}
\hline
\textbf{Reference} & \textbf{Detection Methodology} & \textbf{Model Type} & \textbf{Continous Learning} & \textbf{Handling Unknown Attacks} \\ \hline

\cite{basumallik2019packet} & CNN-based data filter for FDIA in power systems & CNN & No & Not applicable for unknown attacks \\ \hline

\cite{bocu2022real} & Real-time detection and prevention in 5G networks & CNN & No & Low false negative rate limits adaptability to unknowns \\ \hline

\cite{abusitta2019deep} & Proactive intrusion detection using noisy data & Denoising Autoencoder & No & Limited capability for unknown attack handling \\ \hline

\cite{articleSydney} & IDS with feature selection on NSL-KDD dataset & FFDNN & No & Focused on known threats, no handling for new threats \\ \hline

\cite{yin2017deep} & Intrusion detection with binary and multiclass classification & RNN & No & High performance on known threats, limited for unknowns \\ \hline

\cite{10014866} & Lightweight IDS for DDoS attacks in IoT networks &LSTM & No & Unknown attacks are not addressed \\ \hline

\cite{alaei2017incremental} & Incremental detection with Naive Bayes for packet flows & Naive Bayes & Yes & Limited due to data independence assumption \\ \hline

\cite{data2021t} & Learning new knowledge in IDS with DNN & DNN & Partially, but with memory growth issues & Partial handling of unknown attacks \\ \hline

\cite{rebuffi2017icarl} & iCaRL model for catastrophic forgetting & iCaRL  & Yes, with stable memory management & Effective at incremental detection, but requires recalibration \\ \hline

\cite{adjewa2024efficient} & LLMs for traffic analysis in resource-constrained devices & BERT-based & No, static model & Limited as BERT classifies unknown attacks based on known patterns \\ \hline

\cite{Diaf} & Intrusion prediction with GPT and BERT in IoT traffic & LSTM with fine-tuned GPT/BERT & Limited to binary prediction & Effective but lacks detailed analysis for unknown threats \\ \hline

\textbf{Proposed Framework} & \textbf{Continuous detection of emerging attacks using transformer encoder} &\textbf{ BERT-based with Gaussian Mixture Model} & \textbf{Yes, with dynamic clustering} &\textbf{ High adaptability, maintains accuracy with unknown attacks} \\ \hline

\end{tabular}
\end{table*}

\section{Proposed framework}\label{sec:framweork}

\begin{figure*}[ht!]
    \centering
    \includegraphics[width=.7\linewidth]{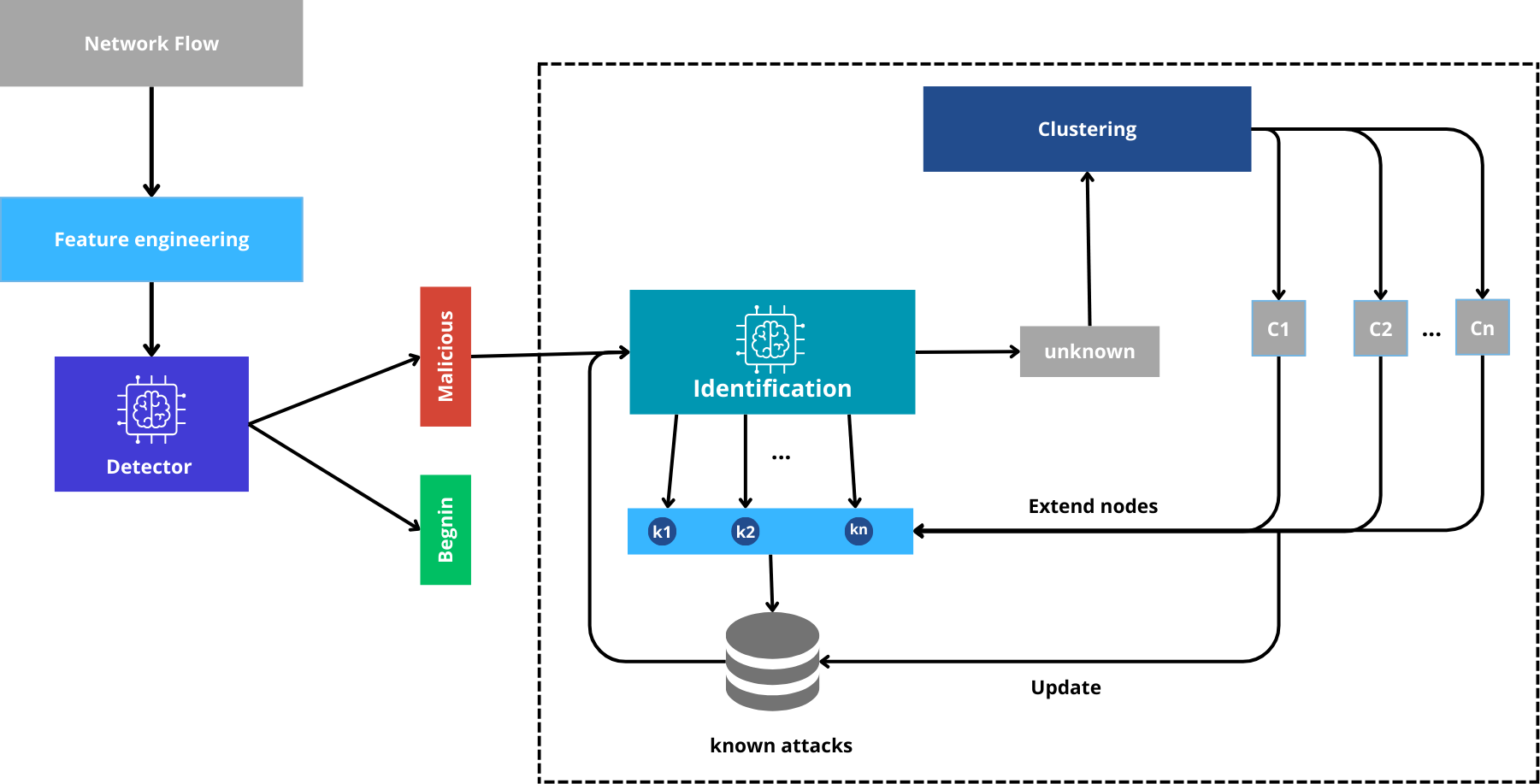}
    \caption{LLM-based Continuous Intrusion Detection Framework Architecture}
    \label{fig:framework}
\end{figure*}

The proposed framework, illustrated in Figure \ref{fig:framework}, consists of several components. First, data are collected from the environment and aggregated for further processing. Then, feature engineering is applied to ensure the data is usable within the system. Since we use a language model to learn patterns within the data, these features are refined \cite{adjewa2024efficient} to optimize model interpretability.

Next, the dataset is divided into training and testing sets to train a binary detection model aimed at distinguishing between malicious and benign traffic. Malicious traffic then continues through the pipeline, together with a small portion of benign traffic to prevent the system from misclassifying unknown  attacks within known attacks that share similar attributes. Within this flow of malicious data, the objective is to identify specific attack types and, additionally, to detect any new or previously unknown attacks.

The identification block, similar to the detection block, uses a language model as its backbone, which is detailed in the following section. We added a softmax classification layer with a fixed number of nodes to accommodate the initial threats we assume are known. This identifier then analyzes the malicious traffic based on this model.
When unknown attacks are detected, they are collected and processed to identify potential hidden clusters. The total number of new clusters, each representing a potential new type of attack, is used to increment the number of nodes in the classification layer. In this way, the model is continuously improved.

\section{\textsc{methodology}}
In this section we present some popular datasets used in IDS and focus on CSE-CIC-IDS2018, we then expose the preprocessing methods we employed, and the design of the model we used as the baseline for our framework. 

\subsection{Datasets and preprocessing}

Numerous datasets have been utilized for intrusion detection systems. \cite{ferrag2020deep} classified these datasets into seven categories. Notable examples include the NSL-KDD dataset, an improved version of KDD'99 addressing redundancy and imbalance \cite{tavallaee2009detailed}, and UNSW-NB15, which features various attacks like DoS and worms across 49 attributes and 2.5 million records \cite{Moustafa}. The CICIDS2017 dataset consists of 80 network flow features and includes attacks like SSH and DDoS, while the CSE-CIC-IDS2018 dataset contains seven attack scenarios, such as DDoS and web attacks \cite{Sharafaldin2018TowardGA}. 
Our framework is evaluated using the CSE-CIC-IDS2018 dataset. However, since it is not compatible with LLMs, we employed the Privacy-Preserving Fixed-Length Encoding (PPFLE) method, which encodes the data for language processing models while preserving privacy using hashing techniques. The ByteLevelBPETokenizer is then used to segment the text into subwords \cite{adjewa2024efficient}.
The original dataset contains approximately 16 million records, but for our experiment, we reduced its size due to computational constraints. We randomly selected 15\% of the benign traffic, given its predominance, and then extracted 30\% of the total dataset while maintaining class proportionality. Critical attack classes were prioritized. Table \ref{tab:attack_counts} presents the final counts for each class.

\begin{table}[ht!]
    \centering
    \caption{Counts of Attack Types}
    \begin{tabular}{@{}lr@{}}
        \toprule
        Class                           & Records    \\ \midrule
        Benign                          & 2022706  \\
        DDoS attack-HOIC               & 205804   \\
        DDoS attacks-LOIC-HTTP         & 172857   \\
        DoS attacks-Hulk               & 138574   \\
        Bot                             & 85857    \\
        FTP-BruteForce                 & 58008    \\
        SSH-Bruteforce                 & 56277    \\
        DoS attacks-SlowHTTPTest       & 41967    \\ \bottomrule
    \end{tabular}
    \label{tab:attack_counts}
\end{table}

\subsection{Model design}
\begin{figure}[H]
    \centering
    \includegraphics[width=\linewidth]{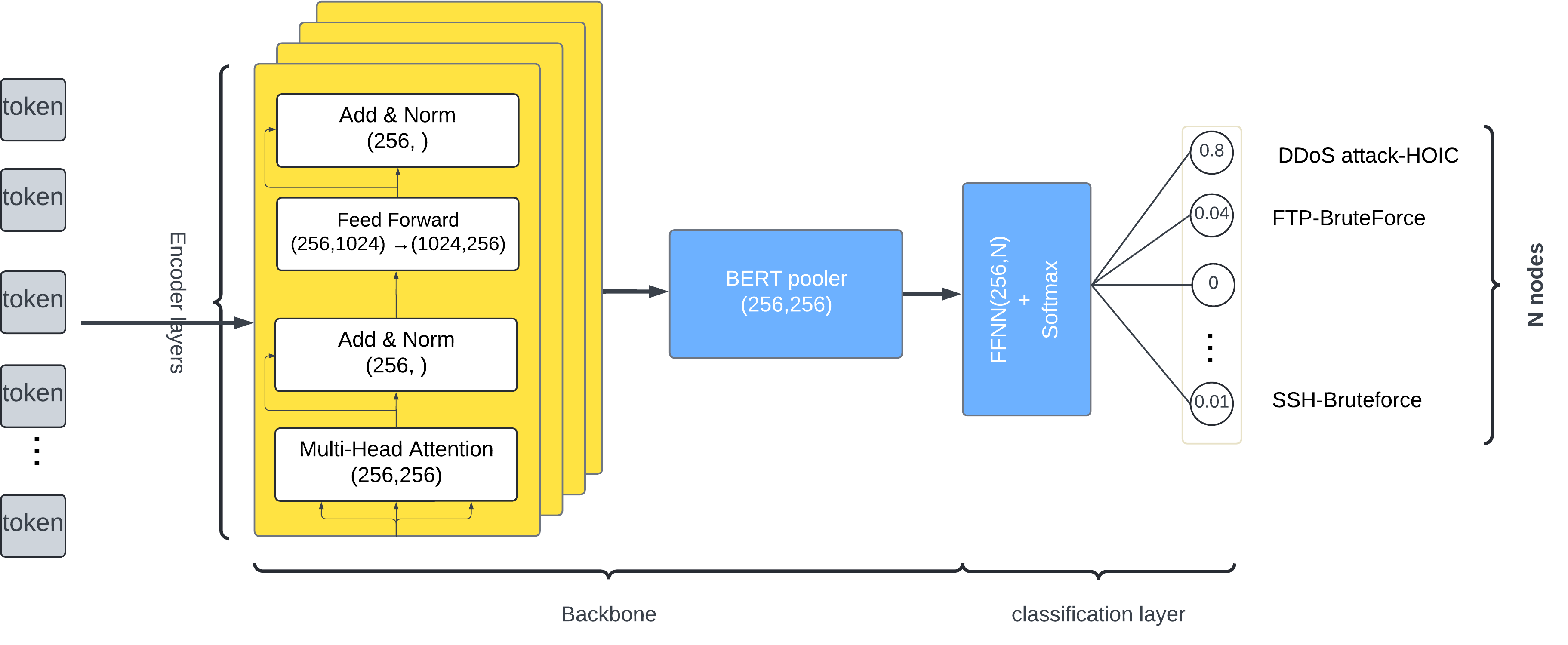} 
    \caption{Model architecture with light BERT as backbone}
    \label{fig:model}
\end{figure}
LLMs are commonly used for tasks such as sentiment analysis, translation, and question answering. However, recent research has demonstrated their potential in detecting network attacks. To leverage this capability, we fine-tuned a BERT model, focusing on its initial layers to efficiently capture the syntactic structure of network flow data.

Specifically, as illustrated in Figure \ref{fig:model}, we retained the first four layers of BERT approximately 10\% of the original model which are primarily responsible for understanding the grammar and syntax of the input data. This approach significantly reduced the model's complexity and computational cost, making it suitable for our purposes.

After processing the network flow data through these layers, a softmax function was applied to generate a probability distribution over different attack classes. This enables the model to classify network flows into specific attack categories with high confidence.

\section{EXPERIMENTATIONS AND DISCUSSION}\label{sec:exp_disc}

To evaluate our framework, we simulated a pipeline to manage the entire process. Firstly, we assessed the detection performance on both benign and malicious traffic. Next, we evaluated the identification performance on the detected malicious traffic, where the framework attempts to identify unknown attacks within specific chunks assumed to contain unknown threats. Chunk 1 included DoS attacks-Hulk and DDoS attack-HOIC, while chunk 2 consisted of SSH brute-force, FTP brute-force, and infiltration attacks.

\subsection{Evaluation metrics}

The metrics employed to assess our approach included \textit{accuracy}. However in the continous learning process, the attack classes are quite imbalanced, as the number of new attacks is unknown. In such scenarios, relying on accuracy can lead to misleading conclusions. Therefore, \textit{precision}, \textit{recall}, and \textit{F1-Score} are utilized to provide a more comprehensive understanding of the model's performance.
\\

\begin{itemize}
    \item \textbf{Accuracy}: Measures the proportion of correct predictions made by the model.
    \[
    Acc = \frac{TP + TN}{TP + TN + FP + FN}
    \]
    
    \item \textbf{Precision}: Quantifies the proportion of correctly predicted positive instances out of all instances predicted as positive.
    \[
    Precision = \frac{TP}{TP + FP}
    \]
    
    \item \textbf{Recall}: Measures the proportion of actual positive instances correctly identified by the model.
    \[
    Recall = \frac{TP}{TP + FN}
    \]
    
    \item \textbf{F1-Score}: The harmonic mean of precision and recall, providing a balanced measure of the model's performance.
    \[
    F1\text{-}Score = 2 \cdot \frac{Precision \cdot Recall}{Precision + Recall}
    \]
\end{itemize}

Where \textit{TP} represents instances of a given class correctly classified as that class, \textit{TN} refers to instances of all other classes correctly classified as not belonging to the given class, \textit{FP} denotes instances incorrectly classified as the given class, and \textit{FN} represents instances of the given class incorrectly classified as belonging to another class.

\subsection{Simulation setup}

For the simulations, we used a Python environment running version 3.12, along with PyTorch 2.0 as the deep learning framework. The computational resources included an L4 GPU with 24 GB of VRAM, which was essential for efficiently handling the computational demands of training the model and processing large datasets.
The use of GPU acceleration allowed us to perform both forward and backward passes faster during model training, significantly reducing the overall computation time, especially when dealing with complex architectures such as those involving LLMs.

\subsection{Results and discussion}

\begin{figure}[ht!]
    \centering
     \includegraphics[width=\linewidth]{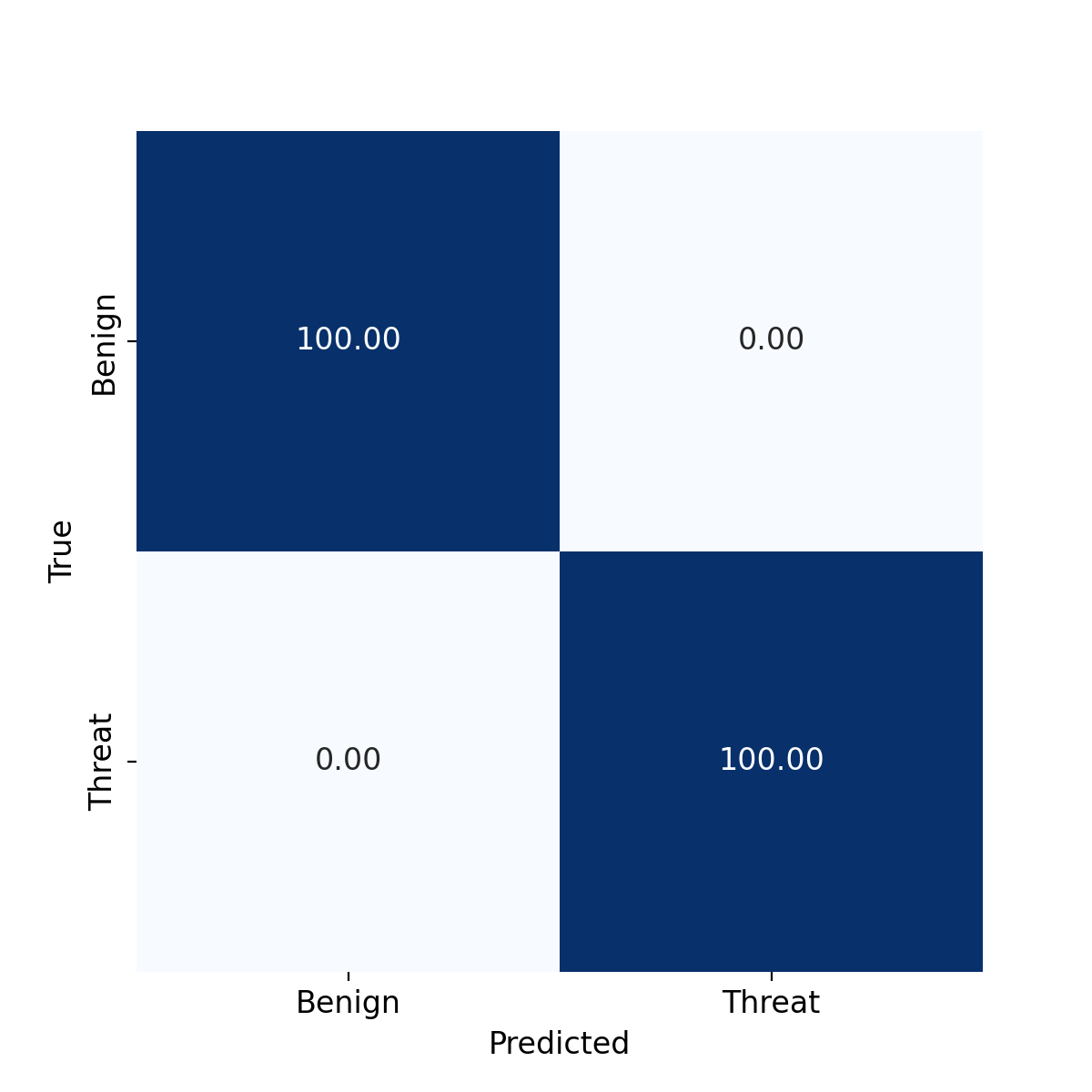}
        \caption{Performance of the Binary Detector}
    \label{fig:bin_clf}
\end{figure}

\begin{figure}[ht!]
    \centering
   \includegraphics[width=\linewidth]{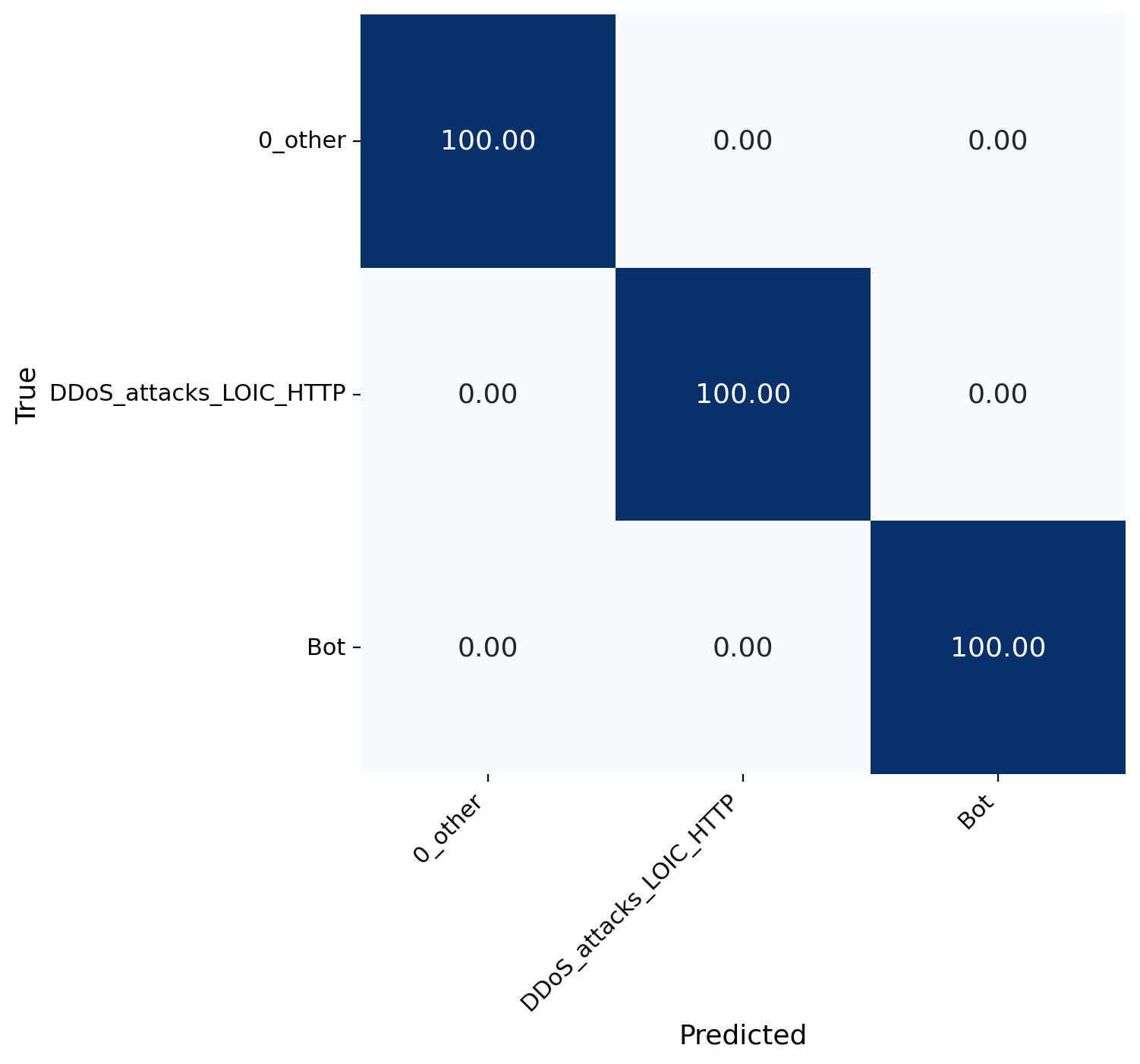}
        \caption{Initial performance of the contious learning model}
    \label{fig:initial}
\end{figure}

\begin{figure}[ht!]
    \centering
    \begin{subfigure}{0.3\textwidth}
        \centering
        \includegraphics[width=\textwidth]{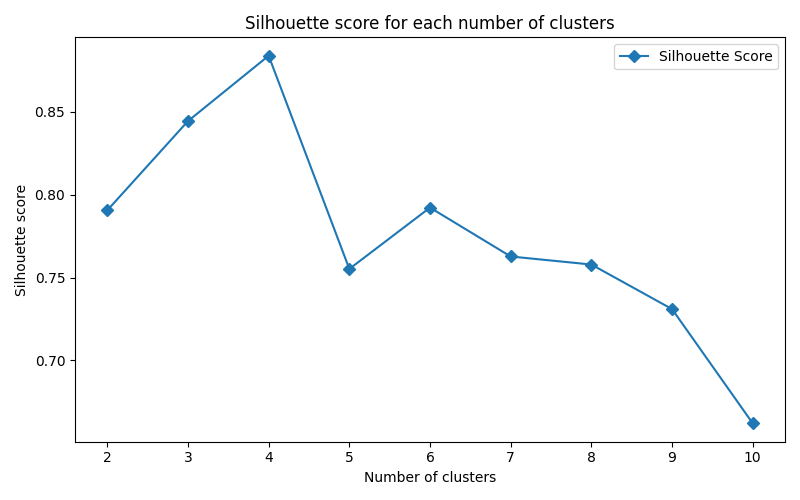}
        \caption{Silhouette score for optimal cluster selection of Chunk 1}
        \label{fig:silhouette1}
    \end{subfigure}
    \hfill
    \begin{subfigure}{0.3\textwidth}
        \centering
        \includegraphics[width=\textwidth]{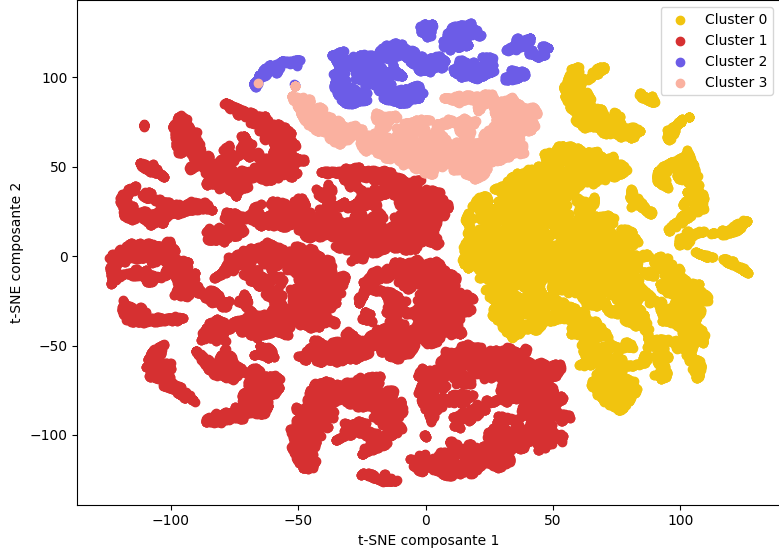}
        \caption{t-SNE 2D visualization of Chunk 1 embeddings}
        \label{fig:tsne_chunk1}
    \end{subfigure}
    \hfill
    \begin{subfigure}{0.3\textwidth}
        \centering
        \includegraphics[width=\textwidth]{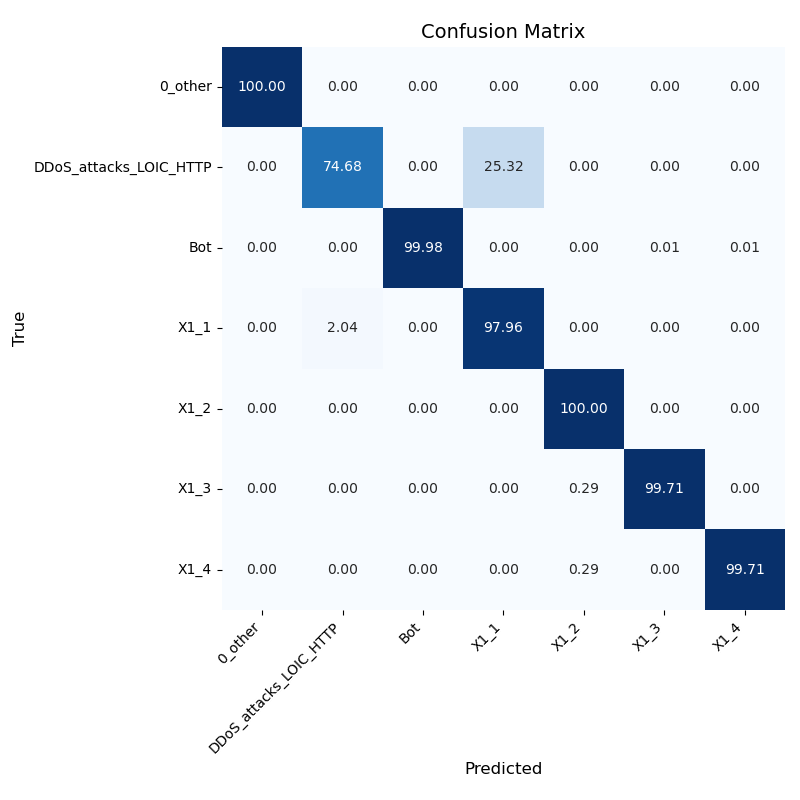}
        \caption{Confusion matrix after adding clusters from Chunk 1}
        \label{fig:cm_chunk1}
    \end{subfigure}
    \caption{Evaluation of clustering results for Chunk 1: silhouette score, t-SNE visualization, and confusion matrix}
    \label{fig:silhouette_analysis}
\end{figure}

\begin{figure}[ht!]
    \centering
    \begin{subfigure}{0.3\textwidth}
        \centering
        \includegraphics[width=\textwidth]{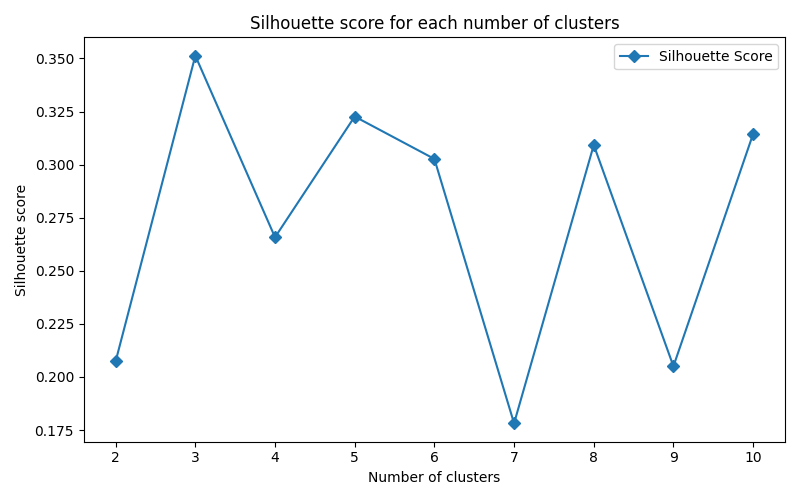}
        \caption{Silhouette score for optimal cluster selection of Chunk 2}
        \label{fig:silhouette2}
    \end{subfigure}
    \hfill
    \begin{subfigure}{0.3\textwidth}
        \centering
        \includegraphics[width=\textwidth]{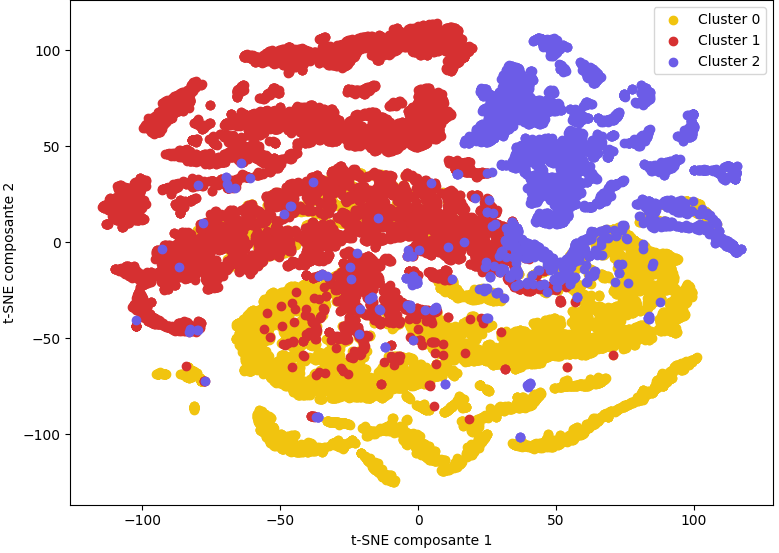}
        \caption{t-SNE 2D visualization of Chunk 2 embeddings}
        \label{fig:tsne_chunk2}
    \end{subfigure}
    \hfill
    \begin{subfigure}{0.3\textwidth}
        \centering
        \includegraphics[width=\textwidth]{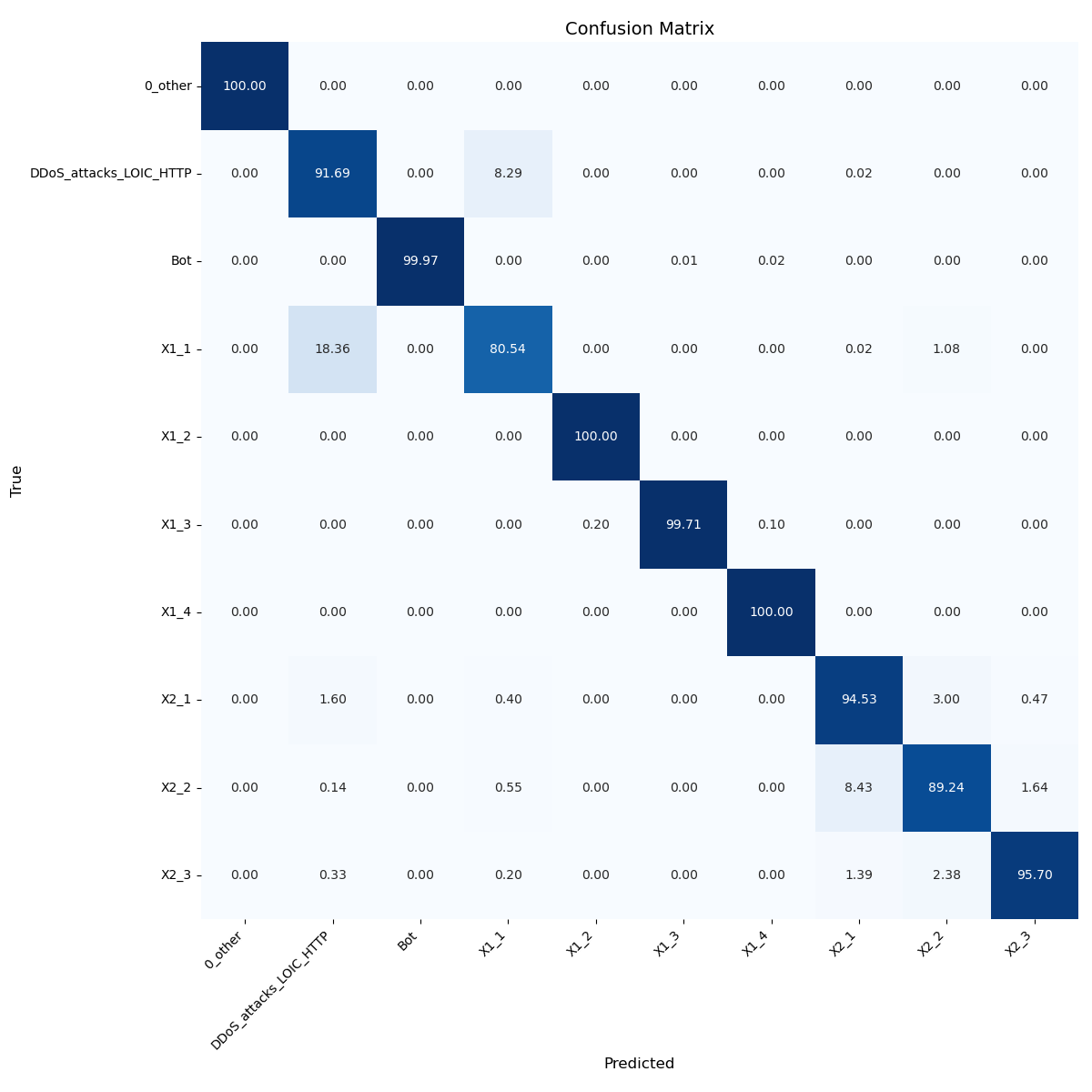}
        \caption{Confusion matrix after adding clusters from Chunk 2 to the updated model}
        \label{fig:cm_chunk2}
    \end{subfigure}
    \caption{Evaluation of clustering results for Chunk 2 on the updated model: silhouette score, t-SNE visualization, and confusion matrix}
    \label{fig:cm_chunks}
\end{figure}

\begin{figure*}[ht!]
    \centering
    \includegraphics[width=\linewidth]{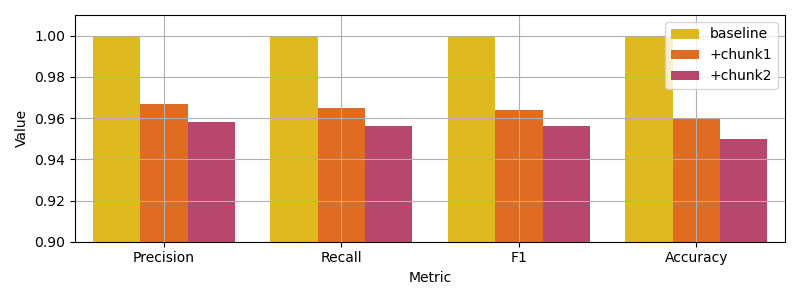}
    \caption{Impact of new threats on Detection Metrics}
    \label{fig:model_performance}
\end{figure*}

The evaluation process starts with the binary detection model, as illustrated in Figure \ref{fig:bin_clf}, it achieves a perfect classification, evidenced by the confusion matrix showing no false positives or false negatives. This indicates that the detector can reliably identify malicious traffic, achieving a recall of 100\%. This performance is critical for ensuring the security of network systems, as it signifies that the model does not misclassify harmful activities as benign, which could lead to significant security breaches.

Next, as shown in Figure \ref{fig:initial}, we assess the identification model using known attack patterns, specifically targeting DDoS attacks-LOIC-HTTP and Bot. During this evaluation, the model effectively identifies these known attacks, demonstrating its robustness. The class "0\_other" plays a crucial role in establishing the decision boundary between known and unknown traffic, allowing the model to distinguish between familiar and unfamiliar patterns.

Subsequently, we introduce two batches of unknown attack traffic to rigorously test the model's ability to detect new or unseen attacks. Chunk 1 contains DoS attacks-Hulk and DDoS attack-HOIC, while chunk 2 consists of SSH brute-force, FTP brute-force, and infiltration attacks. To manage the complexity posed by these unknown attacks, we leverage the Gaussian Mixture Model. Given that the data features are represented as BERT embeddings in high-dimensional space, this probabilistic approach efficiently estimates the likelihood of a data point belonging to a specific cluster. This operation is performed once until new unknown traffic is discovered, ensuring the model adapts to evolving threat landscapes.

The following results provide insights into the model's ability to continuously learn and accurately identify and classify attack patterns over time. In Figure \ref{fig:silhouette1}, the silhouette score for chunk 1 indicates that the optimal number of clusters is 4. This suggests that the traffic consisting of DoS attacks-Hulk and DDoS attacks-HOIC exhibits sufficient diversity to be grouped into four distinct patterns. After clustering, the classification layer is extended with four additional nodes to accommodate the new clusters, enabling the model to update its structure accordingly. This step is crucial for maintaining the model's flexibility. The t-SNE 2D visualization in Figure \ref{fig:tsne_chunk1} further supports this clustering result by projecting the high-dimensional data into two dimensions. As shown, the clusters are well-separated in the 2D space, confirming the effectiveness of the chosen number of clusters. Figure \ref{fig:cm_chunk1} illustrates the impact of adding the newly discovered clusters from chunk 1 to the updated baseline model. A decrease in accuracy is observed, particularly for DDoS attacks-HTTP, which shifts towards the first cluster $X1\_1$. This can be attributed to the overlapping features within the feature space, where shared characteristics between attack types lead to minor misclassifications. Nevertheless, the classification model continues to perform robustly, retaining high accuracy for most attack patterns.

In contrast, the clustering results for chunk 2, shown in Figure \ref{fig:silhouette2}, suggest that the optimal number of clusters is 3. This corresponds to the distinct behaviors of the various attack types in this chunk. As with chunk 1, the classification layer is again extended after clustering to update the model with the newly identified clusters. The t-SNE visualization in Figure \ref{fig:tsne_chunk2} presents the clusters from chunk 2, where the separation is clear. Some blue points, which represent a particular attack type, appear to overlap with the red and yellow points in the 2D plot, but this does not imply misclassification. Rather, this overlap is resolved when viewed in three dimensions, where these clusters are better separated, reinforcing the robustness of the clustering method. Figure \ref{fig:cm_chunk2} shows the model's performance after incorporating the clusters from chunk 2. We observe an improvement in accuracy for DDoS-HTTP attacks, indicating the model's enhanced ability to differentiate between attack types after retraining. This improvement highlights the model's capacity to adapt to new data, as retraining allows it to update its decision boundaries and refine its classifications based on recent information. However, a slight accuracy degradation is noted for the newly learned clusters, such as $X2\_1$ and $X2\_2$, where a small misclassification submatrix appears. This highlights the challenge of maintaining the same level of precision when new attacks are introduced.

To assess the model's ability to evolve over time in the face of unknown attacks, we use the previously mentioned metrics applied to three simulated use cases: (i) The baseline, where the IDS is familiar with all attack types; (ii) the baseline with chunk 1, where the IDS encounters unknown attacks and uses clustering to segment latent unknown attack types, and updates the model, and (iii) the baseline with chunk 1 and chunk 2, where the model has already been updated with attacks from chunk 1 and faces new attacks in chunk 2. In each case, identification, classification, and model updates follow the same steps as for chunk 1. Figure \ref{fig:model_performance} illustrates the impact of exposing the Intrusion Detection System (IDS) to new attacks across various detection metrics, emphasizing the evolution of these metrics as the system encounters previously unseen threats. The "baseline" represents the model's performance on known attacks. The introduction of chunk 1 is followed by an evaluation of identification and classification, which is then repeated with chunk 2. The baseline demonstrates strong performance across all metrics, indicating that the model is well-equipped to classify known attacks.
Upon introducing chunk 1, a slight drop in performance is observed across all metrics, highlighting the challenges of integrating previously unseen attacks. Despite this decrease, the performance remains high due to the effectiveness of the clustering technique, which successfully captures the variations within chunk 1. A further decrease in performance is observed with the addition of chunk 2, as expected, given that this batch includes more complex and diverse attack types. Nonetheless, the overall performance in detecting both known and newly added attacks remains high, showcasing the model's resilience and its ability to maintain security against evolving threats.

\section{CONCLUSION}
In this article, we propose an adaptive framework designed to detect, identify and classify emerging next-generation network attacks. The framework employs a transformer encoder architecture to capture intricate bidirectional patterns in network traffic, achieving strong performance in distinguishing attacks from legitimate activity. Once malicious traffic is detected, the focus shifts to the continuous identification of diverse attack types within this traffic. Leveraging the Gaussian Mixture Model (GMM), the framework effectively clusters features from unknown traffic, enabling dynamic updates to the identification module. Even after integrating new clusters, the framework sustains high detection accuracy, with classification accuracy and recall rates maintaining at 95.6\%. Notably, the model preserves knowledge of previously learned classes, enhancing its capability to differentiate known threats from new ones.

Future research will explore efficient model update techniques that do not rely on prior data, further improving adaptability and performance in dynamic environments.

\section*{Acknowledgement}

We would like to express our gratitude to Saint Jean Ingenieur, a Cameroonian Private Institution of Higher Education, for their generous funding and support, which made this research possible.

\bibliographystyle{plain}
\bibliography{bibliography.bib}

\end{document}